\documentstyle[12pt,a4,amstex,epsfig]{article}
\begin{document}

\begin{flushright}
MPI-PhT/98-76\\
September 1998\\
\end{flushright}
\begin{center}
\large {\bf Complete Helicity Decomposition of the $Bt\bar t$ 
Vertex including Higher Order QCD Corrections and Applications 
to $e^+e^-\rightarrow t \bar t$}\\
\mbox{ }\\
\normalsize
\vskip4cm
{\bf Bodo Lampe}               
\vskip0.3cm
Max Planck Institut f\"ur Physik \\
F\"ohringer Ring 6, D-80805 M\"unchen \\
\vspace{3cm}

{\bf Abstract}\\
\end{center}
The complete density matrix for all polarization 
configurations in the process $B^\ast \rightarrow t\bar t$, 
where $B^\ast$ is an off--shell $Z$ or photon and $t$ is the top quark,
is calculated numerically including oneloop QCD corrections, i.e.  
virtual {\it and} real gluon contributions in $O(\alpha_s)$.  
The analysis is done in the framework 
of the helicity formalism. The results are particularly suited 
for top quark production at the Linear Collider, but may 
be useful in other circumstances as well. 
Relations to LEP and Tevatron physics are pointed out.

\newpage

{\bf \large 1. Introduction} 

Since its discovery in 1995 the top quark has been an object  
of increasing interest. The production process for top quarks 
has been analyzed in various theoretical studies both 
for $pp$ and $e^+e^-$ collisions. Early references on 
the lowest order cross section are \cite{combridge} for 
$pp$ collsions and \cite{reiter} for $e^+e^-$ annihilation. 
Higher order corrections 
to the cross section (total cross section, $p_T$ distribution etc.) 
have been calculated by several groups, \cite{ellis,neerven} 
for $pp$ and \cite{zerwas,jadach} for $e^+e^-$ processes. 
These total cross sections do not involve information 
on the top quark polarization. Such spin effects only come 
in, if one studies distributions of top quark decay products. 
In some cases, spin effects have been studied, i.e. the 
distribution of the top quark spin vector, in ref. 
\cite{parke} for $pp$ and in ref. \cite{arens} for 
$e^+e^-$ collisions, but 
have not been extended to higher orders. 
An interesting step in this direction has been taken in  
ref. \cite{jadach}, where a Monte Carlo 
progam including final state spin terms has been written. 
That article is in fact 
concerned with higher order corrections to $\tau^+\tau^-$ production 
in $e^+e^-$ annihilation at lower energies and so does not take into account 
axial vector couplings. More general results, including 
axial vector couplings, can be found in \cite{groote}, 
but not in the form of a Monte Carlo program. 
Ref. \cite{groote} is an alternative to the approach 
presented here, although we think that our approach 
is more enlightening, systematic and complete. 

Nothing is known about higher order spin corrections to 
the processes which induce top quark production in 
proton collisions (light quark annihilation $q\bar q \rightarrow 
t\bar t$ and gluon--gluon fusion $gg \rightarrow 
t\bar t$). In contrast to $e^+e^-$ annihilation, 
there is initial state gluon radiation and this makes 
those calculations very difficult. The simplifying feature 
of QCD corrections to 
heavy quark production in $e^+e^-$ annihilation is that 
the problem may be reduced to QCD corrections to the 
vectorboson--heavy quark vertex.   
Recently, a method for calculating QCD corrections to 
this vertex has been developed
for the case of top quark decay 
$t\rightarrow b W$ \cite{jirolampe}, i.e. the $tbW$ vertex. In the present 
article this method will be applied and generalized to 
$e^+e^- \rightarrow B(=Z,\gamma)^\ast   \rightarrow t\bar t$.   
The method is based on the helicity formalism 
and allows to obtain the QCD corrections to the
full spin density matrix of the process in a straightforward 
and economical way. 
Furthermore, the results obtained for top quark decay \cite{jirolampe} 
can be easily combined with the results of this paper 
by using a master formula 
Eq. (\ref{mf}) to obtain QCD corrections for any 
distribution of top quark decay products in $e^+e^-$ 
annihilation one may be interested in.  

Within the Standard Model, all couplings of the top quark to other 
particles are completely fixed by its mass and by a few quantum numbers.  
For example, the coupling of the top quark to gluons is a pure 
vector coupling with strength $g_s$, the coupling to a vector boson 
is given by $v_B\gamma_\mu +a_B\gamma_\mu \gamma_5$ with 
$v_Z={1\over 2}-{4\over 3}s_W^2$, $a_Z={1\over 2}$, 
$v_\gamma={4\over 3}s_Wc_W$, $a_\gamma=0$ etc. 
The couplings to the Z--boson and to the photon are particularly 
important for this article, because the process under consideration 
proceeds with an intermediate off--shell Z or photon where 
$Z^\ast$ and $\gamma^\ast$ arise from the annihilation 
of two massless fermions ($e^+$ and $e^-$).  
As will be seen, this latter fact strongly reduces the number of independent 
helicity amplitudes and makes the results quite intuitive.

\vskip1cm

{\bf \large 2. Helicity Description} 

Helicity amplitudes have been considered in many applications 
of phenomenological importance in high energy physics, like 
jet production \cite{dixon}, nonstandard effects in top quark 
processes \cite{kane}, and many others. The idea is, first to 
separate a given process into simpler subprocesses and then to 
explicitly evaluate all  
the possible spin amplitudes for the subprocesses 
in special Lorentz and Dirac frames. The results can afterwards 
be put together with the help of a master formula (to be given below). 
For example, consider the lowest order helicity amplitudes for top quark 
production in $e^+e^-$ annihilation through a vector boson $B^\ast$ 
with either vector or axialvector couplings $v_B$ or $a_B$ to the top 
quark. They are given by 
\begin{equation}
M_V(H,h,\bar h)=v_B \bar u_{h}(p_t) \gamma_{\mu}
v_{\bar h}(p_{\bar t}) \epsilon^{\mu}_{H}
\label{eq1a}
\end{equation}
\begin{equation}
M_A(H,h,\bar h)=a_B \bar u_{h}(p_t) \gamma_{\mu} \gamma_5
v_{\bar h}(p_{\bar t}) \epsilon^{\mu}_{H}
\label{eq1b}
\end{equation}
where 
$h=\pm {1\over 2}$, $\bar h=\pm {1\over 2}$ and $H=0,\pm 1$ 
label the spin states for the top quark and the 
B--boson. Note that for off-shell photons there is a 
longitudinal component with $H=0$ just as for a massive 
vector boson. 
In total, there are 24 amplitudes to be considered, which can in 
principle be relevant to the process. 

Real gluon emission cannot be treated on the amplitude level, 
but only on the level of the density matrix
\begin{equation}
\rho_{IJ}(H,h,\bar h,H',h',\bar h') :=                
M_I(H,h,\bar h)M_J^\ast (H',h',\bar h')
\label{mf8877}
\end{equation} 
where $I$ and $J$ denote either $V$(=vector) or $A$
(=axialvector). The gluons are supposed to 
be unpolarized. 
After hermiticity 
one has 78 independent density matrix elements 
each for the $VV$, $VA$ and $AA$ case, a total of 234 amplitudes. 
As will be shown below, only 60 of these are relevant for 
$e^+e^-$ annihilation.  
Note that the amplitudes are only determined up to an overall phase, 
and that 
this arbitrariness goes away when forming the density matrix elements.  

The full density matrix is in fact needed in the 'master formula', 
if one considers the combined production and decay process 
for the top quark. Assume the general case, that top quarks are produced 
in some process $ab \rightarrow t\bar t$ and then decay 
according to $t \rightarrow W^+b$ and $\bar t \rightarrow W^-\bar b$, 
where the W's further decay to light (massless) fermions, $ W^+ \rightarrow 
f_1 \bar f_2$ and $ W^- \rightarrow 
f_3 \bar f_4$. The cross section is then given by the 'master formula' 
\begin{equation}
\sigma =\sum_{EXT} \vert \sum_{INT} M(h_a,h_b,h,\bar h)
M(h,h_{W^+})M(\bar h,h_{W^-})M(h_{W^+})M(h_{W^-})\vert ^2
\label{mf}
\end{equation}
where $EXT=h_a,h_b$ denotes the spins of the external 
particles and $INT=h,\bar h,h_{W^+},h_{W^-}$ the spins of 
the internal particles of the process. 
$M(h_a,h_b,h,\bar h)$ are the helicity amplitudes for the 
production process, $M(\bar h,h_{W^-})$ for the decay of the 
antitop quark, and $M(h_{W^+})$ and $M(h_{W^-})$ are the amplitudes 
for the decay of the $W^+$ and $W^-$, respectively. 
The fact that
the formula makes no explicit reference to the spins of the 
massless fermions $f_i$, i=1,2,3,4, has the same reason as 
the non--appearance of the b--quark spins. Namely, the $h_{f_i}$ are 
fixed by the V--A nature of the W decays, just as $h_b$ is fixed 
by the V--A nature of the top decays (assuming massless b--quarks). 
Furthermore, if one is not interested in the decay of the antitop 
or of the W's, the corresponding amplitudes and helicities will 
not appear in  
the above formulas. In that case, the $\bar t$ and/or the W's 
will be one of the external (EXT) particles, whose spins 
have to be summed over after taking the square in Eq. (\ref{mf}).    
Besides neglecting the b--quark mass, it is also a good 
approximation in Eq. (\ref{mf}) to take the internal particles 
(top quarks and W's) 
on--shell, because off-shell contributions are suppressed by powers 
of the width $\Gamma_t$ and $\Gamma_W$. 

\vskip1cm

{\bf \large 3. Deconstruction of the $e^+e^-$ Production Cross Section}

In contrast, the intermediate vector boson $Z$ and $\gamma$ 
must not be taken on--shell. The general spin amplitude 
for the production part of the process is
\begin{equation}
M(h_p,h_e,h,\bar h)={M_Z(h_p,h_e,h,\bar h)\over Q^2-m_Z^2}
    +{M_\gamma(h_p,h_e,h,\bar h)\over Q^2}
\label{mgz}
\end{equation}
where $h_{p,e}$ are the $e^\pm$ spins and 
$Q^2$ is the square of the $e^+e^-$ energy. 
These amplitudes are now to be decomposed into the elementary 
amplitudes Eqs. (\ref{eq1a}) and (\ref{eq1b}). 
The decompostion will be on the algebraic level only, and no 
numerical analysis will be presented in this section, because we just 
want to show how the elementary amplitudes enter the 
$e^+e^-$ cross section. Afterwards the complete 
oneloop QCD corrections to all the elementary amplitudes 
resp. spin density matrix elements will be given (sections 5 and 6). 

\begin{figure}
\begin{center}
\epsfig{file=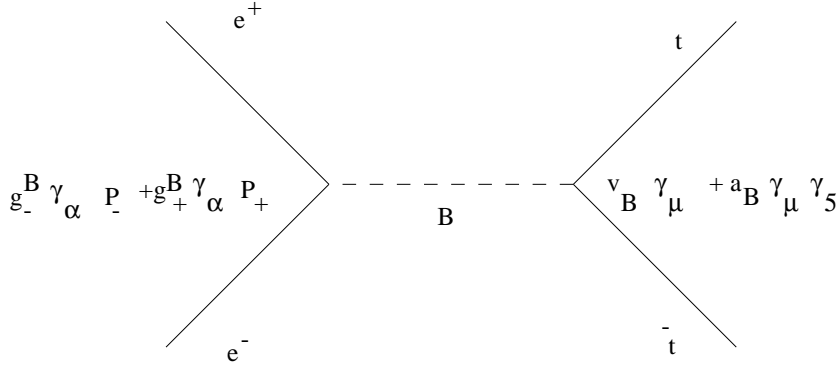,height=11cm,angle=270}
\bigskip
\caption{Definition of the couplings}
\end{center}
\end{figure}

The two amplitudes $M_Z$ and $M_\gamma$ 
in Eq. (\ref{mgz}) have a 
decomposition of the form 
\begin{equation}
M_B(h_p,h_e,h,\bar h)=\sum_H M_B(h_p,h_e,H) {1\over Q^2-m_B^2} M_B(H,h,\bar h) 
\label{mgzd}
\end{equation}
where the sum is over the Spins $H=0,\pm 1$ of the off-shell B--boson. 
In this way the $Bt\bar t$ vertex amplitudes 
Eqs. (\ref{eq1a}) and (\ref{eq1b})
enter the cross section 
(\ref{mf}). $M_B(h_p,h_e,H)$ are the spin amplitudes for 
$e^+e^-\rightarrow B^\ast$ and have a very simple structure due 
to the approximate masslessness of $e^\pm$. Namely, one has 
\begin{equation}
M_B(h_p,h_e,H)=
\begin{cases}
{g_-^B\over \sqrt 2}  &  \text{for} \quad (h_p,h_e,H)=(+{1\over 2},-{1\over 2},-1) \\
{g_+^B\over \sqrt 2}  &  \text{for} \quad (h_p,h_e,H)=(-{1\over 2},+{1\over 2},+1) \\
0                     & \text{otherwise}   
\end{cases} 
\label{cas1}
\end{equation}
where $g_\pm^B=v_e^B\pm a_e^B$ are the left-- and right--handed 
part of the lepton couplings to the $B$, with   
$v_e^Z=-{1\over 2}+2s_W^2$, $a_e^Z=-{1\over 2}$,
$v_e^\gamma=-2s_Wc_W$, $a_e^\gamma=0$.
Furthermore, 
the spin quantization axis has been chosen to be the $+z$ direction.   
According to Eq. (\ref{cas1}) not all of the amplitudes 
$M_B(H,h,\bar h)$ are picked up in Eq. (\ref{mgzd}) to give a 
nonzero contribution to the $e^+e^-$ amplitudes $M_B(h_p,h_e,h,\bar h)$. 
Fig. 1 gives a graphic view about what notations are used 
for the couplings of the lowest order process. From this figure  
it becomes clear that the amplitudes $M_B(H,h,\bar h)$ 
are a linear combination $\sim v_BM_V+a_BM_A$. Altogether one gets:  
\begin{equation}
M_B(h_p,h_e,h,\bar h)=
\begin{cases}
+{g_-^B\over \sqrt 2}[v_BM_V(-1,h,\bar h)+a_BM_A(-1,h,\bar h)] &  
                                  h_p=+{1\over 2}, h_e=-{1\over 2}   \\
-{g_+^B\over \sqrt 2}[v_BM_V(+1,h,\bar h)+a_BM_A(+1,h,\bar h)] &  
                                  h_p=-{1\over 2}, h_e=+{1\over 2}   \\
0                     & \text{otherwise}
\end{cases} 
\label{mgzd59}
\end{equation}
One can now combine the contributions from $\gamma$ and $Z$ 
(cf. Eq. (\ref{mgz})):
\begin{equation}
M(h_p,h_e,h,\bar h)=
\begin{cases}
v_-M_V(-1,h,\bar h)+a_-M_A(-1,h,\bar h)& 
                                  h_p=+{1\over 2}, h_e=-{1\over 2}   \\
v_+M_V(+1,h,\bar h)+a_+M_A(+1,h,\bar h)& 
                                  h_p=-{1\over 2}, h_e=+{1\over 2}   \\
0                     & \text{otherwise}
\end{cases}
\label{mgzd60}
\end{equation}
where we have used the abbreviations 
$v_\pm=\mp {1\over \sqrt 2} 
      ({g_\pm^Zv_Z\over Q^2-m_Z^2}+{g_\pm^\gamma v_\gamma\over Q^2})$ 
and
$a_\pm=\mp {1\over \sqrt 2} 
      {g_\pm^Za_Z\over Q^2-m_Z^2}$ . 
According to Eqs. (\ref{mf}) and (\ref{mgzd60}), 
if the $e^\pm$ beams are unpolarized, the cross section is  
a sum of two terms, $\sigma=\sigma_-+\sigma_+$, where the first 
term is due to $e^+e^-$ helicities 
$h_p=+{1\over 2}$, $h_e=-{1\over 2}$ and the second term 
is due to $e^+e^-$ helicities $h_p=-{1\over 2}$, $h_e=+{1\over 2}$. 
One has 
\begin{equation}
\sigma_\pm =\vert \sum_{h,\bar h} 
[v_\pm M_V(\pm 1,h,\bar h)+a_\pm M_A(\pm 1,h,\bar h)] 
               D(h,\bar h) \vert^2
\label{mgzd61}
\end{equation}
where the contributions from the decay amplitudes have been 
summarized as $D(h,\bar h)$. 
When one carries out the modulus squared in 
Eq. (\ref{mgzd61}), it becomes apparent that in general all the 
density matrix elements of the form 
$M_I(+1,h,\bar h)M_J(+1,h',\bar h')^{\ast}$ and 
$M_I(-1,h,\bar h)M_J(-1,h',\bar h')^{\ast}$ 
for $IJ=VV,VA+AV$ and $AA$ are needed 
to calculate the cross section of the decay products.  
Note that instead of $VA$ we are using the explicitly 
hermitean combination $VA+AV$. This is no restriction 
because to all orders in QCD one has $M_VM_A^{\ast}=M_AM_V^{\ast}$. 
As will become explicit in section 5, the matrix 
elements of $H=+1$ and $H=-1$ are related, so that 
only the first set $M_I(+1,h,\bar h)M_J(+1,h',\bar h')^{\ast}$ 
has to be calculated. 
For each combination $IJ=VV,VA+AV$ and $AA$ 
there are 10 independent matrix elements in this set.

\vskip1cm

{\bf \large 4. The Method}

The results to be presented were obtained in the helicity 
formalism. Furthermore, they concern the oneloop QCD corrections 
to the lowest order matrix elements. The lowest order expressions 
are usually simple, whereas the oneloop expressions, in particular 
for the case of hard gluons are quite lenghty for arbitrary 
spin orientations of $B$, $t$ and $\bar t$ (the gluons are 
assumed to be unpolarized) 
\footnote{see however ref. \cite{schmi} for rather compact 
formulae}. 
Furthermore, the hard gluon contributions 
cannot be treated on the level of amplitudes, because they 
have to be integrated over the gluon's energy and angles. 
One really has to go to the level of the (spin) density 
matrix, 
\begin{equation}
\rho (H,h,\bar h,H',h',\bar h'):=M(H,h,\bar h)M^\ast(H',h',\bar h')
\label{dm25z}
\end{equation} 
to do the phase space 
integrations. There is, however, one circumstance which simplifies 
the task. This is related to the fact, that the oneloop 
QCD corrections to the total (spin--averaged) cross section for 
top quark production are known \cite{zerwas,report}. 
This allows to get 
rid of the infrared and collinear singularities present 
in the matrix elements, by forming suitable singularity--free 
combinations of the spin--dependent and the spin--averaged 
expressions. The point is that the infrared and collinear 
singularities are 'universal', i.e. independent of the spin 
direction, so that they will drop out in suitable ratios and 
differences. 
To be more explicit, consider the ratio 
\begin{equation}
R(H,h,\bar h,H',h',\bar h') := {\rho (H,h,\bar h,H',h',\bar h')\over 
\text{tr} \rho}
\label{dm25}
\end{equation}                      
where the trace of the density matrix is given by 
\begin{equation}
\text{tr} \rho =\sum_{H,h,\bar h} 
\vert M(H,h,\bar h) \vert^2
\label{dm250}
\end{equation}
Now assume that each matrix element has the generic form 
$\rho =T+\alpha K$ where $T$ stands for the tree level term,  
$K$ for the higher order correction and $\alpha=C_F{\alpha_s\over 2\pi}$ 
is the QCD coupling constant and $C_F={4\over 3}$. Assume similarly, that 
$\text{tr} \rho$ has the form 
$\text{tr} \rho =T_{\text{all}}+\alpha K_{\text{all}}$. Then 
the ratio $\rho / \text{tr} \rho$ is given by 
\begin{equation}
R= {\rho \over \text{tr} \rho} =
{T\over T_{\text{all}}} 
+\alpha {KT_{\text{all}}-TK_{\text{all}}\over T^2_{\text{all}}  }
+O(\alpha^2) =\text{LO} +\alpha \, \text{HO}
\label{dm25aa}
\end{equation}
It turns out that the difference  
$KT_{\text{all}}-TK_{\text{all}}$ is free of infrared 
and collinear (and ultraviolet) singularities. If calculated, 
it gives an elegant means to determine higher order corrections 
to the density matrix \cite{lampe}. 
In the formulas and figures presented below, $R$ will be 
written as $R=\text{LO} +\alpha \, \text{HO}$ where 
LO and HO will be plotted as a function of $m_t/Q$. 

Finally note that the considerations in this section 
should in principle be done for the three matrices 
$\rho_{VV}$, $\rho_{VA}(=\rho_{AV})$ and $\rho_{AA}$
separately. In the case of $VA$ it turns out that 
the trace of the density matrix vanishes (in leading order 
and in higher order). It is therefore 
convenient to consider a different linear combination, e.g. 
$VA+AV+AA$ with a nonvanishing $\text{tr} \rho$, and to apply 
the method Eq. (\ref{dm25aa}) on that. Results for 
$VA$ can afterwards be reconstructed by taking 
the difference of $VA+AV+AA$ and $AA$ (see Eq. (\ref{md610}) in the 
summary section). 

\vskip1cm

{\bf \large 5. Lowest Order Amplitudes and Virtual Corrections}

I have calculated the eight needed amplitudes $M_I(+1,h,\bar h)$, $I=V,A$  
as defined in Eqs. (\ref{eq1a}) and (\ref{eq1b})  
using the 'chiral representation' 
of $\gamma$ matrices, in which $\gamma_5=$diag$(-1,-1,1,1)$ etc. 
The results are presented in the rest frame of the virtual 
B--boson where the momenta are given by 
\begin{equation}
p_B=(Q,0,0,0) \qquad p_t={Q\over 2} (1,0,0,\beta) 
\qquad p_{\bar t}={Q\over 2} (1,0,0,-\beta)
\label{dm28}
\end{equation}
with $\beta^2=1-{4m_t^2\over Q^2}$. An independent set of 
spinors for the top quarks is then given by 
\begin{eqnarray} \nonumber
&\bar u_{+{1\over 2}} (p_t)=(a_+,0,a_-,0) 
\qquad 
&\bar u_{-{1\over 2}} (p_t)=(0,a_-,0,a_+)
\\ \nonumber 
& v_{+{1\over 2}} (p_{\bar t})=(-a_+,0,a_-,0) 
\qquad      
& v_{-{1\over 2}} (p_{\bar t})=(0,-a_-,0,a_+)
\label{q68}
\end{eqnarray}  
where $a_\pm^2={Q\over 2}(1\pm \beta)$. 
The quantization axis of the B--boson spin is chosen to be at an 
angle $\theta$ w.r.t. the top quark momentum, i.e. 
the polarization vector for $H=+1$ is given by 
\begin{equation}
\epsilon_{+ 1}={1\over \sqrt{2}}(0,\cos\theta,-i,-\sin\theta)
 \, .
\label{q5}
\end{equation} 
The most general parametrization of polarization vectors would be 
\begin{eqnarray} \nonumber
\epsilon_{- 1}&=&-{e^{i\phi}\over \sqrt{2}}(0,
\cos\phi\cos\theta -i\sin\phi,\sin\phi\cos\theta+i\cos\phi,-\sin\theta)
\\ \nonumber 
\epsilon_{+1}&=&-\epsilon_{- 1}^\ast 
\\ 
\epsilon_0&=&-(0,\sin\theta\cos\phi,\sin\theta\sin\phi,\cos\theta) \, .
\label{q52}
\end{eqnarray}
but for the problem at hand one may choose $\phi =0$ without 
restriction. Note further that although the off--shell B--boson 
has a longitudinal polarization component, there is no contribution from 
longitudinal $B's$ to the $e^+e^-$ cross section, as has been 
shown in section 3. 
Nevertheless, for completeness and curiosity, all the lowest 
order amplitudes are given here:  
\begin{eqnarray} \nonumber
&M_V(0,-{1\over 2},-{1\over 2})=-2{m_t\over Q}\cos\theta 
\quad
&M_A(0,-{1\over 2},-{1\over 2})=0
\\ \nonumber 
&M_V(-1,-{1\over 2},-{1\over 2})=\sqrt 2 {m_t\over Q} \sin\theta e^{i\phi} 
\quad
&M_A(-1,-{1\over 2},-{1\over 2})=0
\\ \nonumber
&M_V(+1,-{1\over 2},-{1\over 2})=-\sqrt 2 {m_t\over Q}\sin\theta e^{-i\phi} 
\quad
&M_A(+1,-{1\over 2},-{1\over 2})=0
\\ \nonumber
&M_V(0,-{1\over 2},+{1\over 2})=\sin\theta e^{i\phi}
\quad
&M_A(0,-{1\over 2},+{1\over 2})=\beta\sin\theta e^{i\phi}
\\ \nonumber
&M_V(-1,-{1\over 2},+{1\over 2})=-e^{2i\phi}{1-\cos\theta\over\sqrt 2}
\quad
&M_A(-1,-{1\over 2},+{1\over 2})=-\beta e^{2i\phi}{1-\cos\theta\over\sqrt 2}
\\ \nonumber
&M_V(+1,-{1\over 2},+{1\over 2})=-{1+\cos\theta\over\sqrt 2}
\quad
&M_A(+1,-{1\over 2},+{1\over 2})=-\beta{1+\cos\theta\over\sqrt 2}
\\ \nonumber
&M_V(0,+{1\over 2},-{1\over 2})=\sin\theta e^{-i\phi}
\quad
&M_A(0,+{1\over 2},-{1\over 2})=-\beta\sin\theta e^{-i\phi}
\\ \nonumber
&M_V(-1,+{1\over 2},-{1\over 2})={1+\cos\theta\over\sqrt 2}
\quad
&M_A(-1,+{1\over 2},-{1\over 2})=-\beta{1+\cos\theta\over\sqrt 2}
\\ \nonumber
&M_V(+1,+{1\over 2},-{1\over 2})=e^{-2i\phi}{1-\cos\theta\over\sqrt 2}
\quad
&M_A(+1,+{1\over 2},-{1\over 2})=-\beta e^{-2i\phi}{1-\cos\theta\over\sqrt 2}
\\ \nonumber
&M_V(0,+{1\over 2},+{1\over 2})=2{m_t\over Q}\cos\theta
\quad
&M_A(0,+{1\over 2},+{1\over 2})=0
\\ \nonumber
&M_V(-1,+{1\over 2},+{1\over 2})=-\sqrt 2 {m_t\over Q} \sin\theta e^{i\phi}
\quad
&M_A(-1,+{1\over 2},+{1\over 2})=0
\\ 
&M_V(+1,+{1\over 2},+{1\over 2})=\sqrt 2 {m_t\over Q} \sin\theta e^{-i\phi}
\quad
&M_A(+1,+{1\over 2},+{1\over 2})=0
\label{q6ttt}                       
\end{eqnarray}                   
These amplitudes show a lot of symmetry. The most important 
for us arises from interchanging the role of particle and 
antiparticle. For $\phi=0$ this amounts to $\theta \leftrightarrow 
\theta+\pi$ and interchanges amplitudes of the form 
$M_I(+1,h,\bar h)\leftrightarrow M_I(-1,h,\bar h)$, $I=V,A$. 
In fact this relation is induced by CP invariance and it holds 
including higher order QCD corrections.  
It reduces the number of independent density matrix elements 
to be calculated by half from 60 to 30. 

Using Eq. (\ref{q6ttt}) one may calculate the trace of the 
corresponding density matrix Eq. (\ref{dm250}): 
\begin{equation}
\text{tr} \rho_{VV} =4+8{m_t^2\over Q^2} \qquad 
\text{tr} \rho_{VA} =\text{tr} \rho_{AV} =0 \qquad
\text{tr} \rho_{AA} =4\beta^2 
\label{dm251}
\end{equation}
These expressions should take the role of $T_{\text{all}}$ 
in Eq. (\ref{dm25}). However, 
it appears that for the VA interference term 
the ratio $R_{VA}\equiv \rho_{VA} /\text{tr} \rho_{VA}$ becomes infinite. 
Instead of 
$R_{VA}$ the following combination will be considered in the 
following 
$R_{VAAVAA}:=(\rho_{VA}+\rho_{AV}+\rho_{AA})/\text{tr} \rho_{AA}$. 

The next step is to incorporate the corrections from virtual 
gluon exchange. This can be done on the amplitude level and is 
quite straightforward, because virtual gluons do not 
modify the kinematics of the lowest order process. 
The effect can be condensed to effectively change the vector and 
axialvector interactions according to 
\begin{equation}
\gamma_\mu\rightarrow\gamma_\mu(1+\alpha f_1) 
    +{i\over 2m_t} \sigma_{\mu\nu}p_B^\nu\alpha f_2    \qquad \qquad
\gamma_\mu\gamma_5\rightarrow\gamma_\mu\gamma_5(1+\alpha f_A)
\label{dm251x}
\end{equation}
where $\alpha=C_F{\alpha_s\over 2\pi}$ as before and 
$f_1$, $f_2$ and $f_A$ are functions of $\beta$ and 
can be found in \cite{zerwas}. It turns out that $f_2=f_1-f_A$. 
Therefore, 
the amplitudes Eq. (\ref{q6ttt}) can be easily extended 
to contain effects from virtual gluon exchange. We here give 
the result only for the relevant cases $\phi=0$ and $H=+1$: 
\begin{eqnarray} \nonumber
&M_V(+1,-{1\over 2},-{1\over 2})=-\sqrt 2 {m_t\over Q}\sin\theta 
                 (1+\alpha f_+) 
\quad
&M_A(+1,-{1\over 2},-{1\over 2})=0
\\ \nonumber
&M_V(+1,-{1\over 2},+{1\over 2})=-{1+\cos\theta\over\sqrt 2}
                 (1+\alpha f_-) 
\quad
&M_A(+1,-{1\over 2},+{1\over 2})=-\beta{1+\cos\theta\over\sqrt 2}
                 (1+\alpha f_-) 
\\ \nonumber
&M_V(+1,+{1\over 2},-{1\over 2})={1-\cos\theta\over\sqrt 2}
                 (1+\alpha f_-) 
\quad
&M_A(+1,+{1\over 2},-{1\over 2})=-\beta {1-\cos\theta\over\sqrt 2}
                 (1+\alpha f_-) 
\\ 
&M_V(+1,+{1\over 2},+{1\over 2})=\sqrt 2 {m_t\over Q} \sin\theta 
                 (1+\alpha f_+) 
\quad
&M_A(+1,+{1\over 2},+{1\over 2})=0
\label{q677}                       
\end{eqnarray}                   
where $f_+=f_1-{Q^2\over 4m_t^2}f_2$ and $f_-=f_A=f_1-f_2$. 

It should be noted that 
the functions $f_1$ and $f_A$ merely 'renormalize' the form of 
the lowest order interactions, cf. Eq. (\ref{dm251x}), 
and accordingly they {\it do not 
contribute} in the ratios 
$R_{IJ}=\rho_{IJ} / \text{tr} \rho_{IJ}$, $IJ=VV,VAAVAA$ and $AA$. 
The interesting point is  
that the infrared and collinear singularities are solely contained 
in the functions $f_1$ and $f_A$ whereas $f_2$ is completely finite 
and given by 
\begin{equation}
f_2={1-\beta^2\over 2\beta}\ln {1-\beta\over 1+\beta}
\label{dm253}
\end{equation}
The normalized density matrix elements $R_{VV}(+1,h,\bar h,+1,h',\bar h')$ 
are given in the table 
\begin{table}[h] 
\label{tab1}  
\begin{center}
\begin{tabular}{|l|l|l|l|l|}
\hline
$h\bar h\downarrow\vert h'\bar h'\rightarrow$ & 
     $(-{1\over 2},-{1\over 2})$    &  $(-{1\over 2},+{1\over 2})$ &
   $(+{1\over 2},-{1\over 2})$    &  $(+{1\over 2},+{1\over 2})$  \\
\hline
$(-{1\over 2},-{1\over 2})$ & $-4\beta^2s_t^2\alpha f_2$ & 
$\star$ & $\star$ &$\star$    \\
$(-{1\over 2},+{1\over 2})$ & 
$-\lambda\beta^2s_t(1+c_t)\alpha f_2$ &
$2\beta^2(1-c_t)^2\alpha f_2$ & $\star$ &$\star$    \\
$(+{1\over 2},-{1\over 2})$ &
$\lambda\beta^2s_t(1-c_t)\alpha f_2$ &
$-2\beta^2s_t^2\alpha f_2$ &
$2\beta^2(1-c_t)^2\alpha f_2$ &$\star$    \\
$(+{1\over 2},+{1\over 2})$ &
$4\beta^2s_t^2\alpha f_2$ & 
$\lambda\beta^2s_t(1+c_t)\alpha f_2$ &
$-\lambda\beta^2s_t(1-c_t)\alpha f_2$ &
$-4\beta^2s_t^2\alpha f_2$ \\
\hline
\end{tabular}
\bigskip
\end{center}
\end{table}
where $\lambda={1-2m_t^2/Q^2\over m_t/Q}$, 
$s_t=\sin\theta$ and $c_t=\cos\theta$ and the $\star$'s 
follow from the symmetry of the matrix.
It is thereby explicit that all corrections are $\sim f_2$ 
as anticipated.  

The density matrix elements of $R_{AA}$ and 
$R_{VAAVAA}$ do not get any corrections at all from virtual 
gluon exchange because according to Eq. (\ref{q677}) there are 
two factors of $1+\alpha f_-$ both in the numerator and denominator 
of $\rho_{AA}/\text{tr} \rho_{AA}$ and similarly 
for $R_{VAAVAA}$.  

\vskip1cm

{\bf \large 6. Real Gluon Emission and Numerical Results for the Corrections 
to the Normalized Density Matrix}

Next we come to the two Feynman diagrams with real gluon emission 
$B^\ast \rightarrow t\bar t g$. The higher order corrections are 
calculated in such a way that all gluon d.o.f. are summed and 
integrated over. This means, for example, the gluon is assumed 
to be unpolarized. Furthermore, a rather complicated phase space 
integrattion has to be performed. In contrast to the tree level 
process $B^\ast \rightarrow t\bar t$ where the phase space is 
trivial, one has here two highly nontrivial integrations which 
I have choosen to perform numerically. This is then straightforward 
because, as pointed out before, the integrand 
corresponding to Eq. (\ref{dm25aa}) is 
completely finite.  

The trivial phase space for the lowest order kinematics 
(\ref{dm28}) is given by PS$(B^\ast \rightarrow t\bar t)=\int 
\prod {d^3p_i\over 2E_i}\delta^4(p_t+p_{\bar t}-p_B)={\pi\over 2}\beta$. 
With an additional gluon the 4--momenta of the particles become 
more complicated. First, the energies of $t$ and $\bar t$ are not 
fixed as in lowest order -- in fact they are to become the integration 
variables -- and secondly, there is now an angle $\omega$ between 
the top quark and antitop direction. 
\begin{equation}
p_B=(Q,0,0,0) \quad p_t={Q\over 2} x_1(1,0,0,\beta_1)
\quad p_{\bar t}={Q\over 2}x_2 (1,\beta_2\sin\omega,0,\beta_2\cos\omega)
\label{dm2821}
\end{equation}
Here $x_1$ and $x_2$ are the (normalized) energies of $t$ and $\bar t$
and $\beta_i^2:=1-{4m_t^2\over x_i^2Q^2}$, $i=1,2$. 
$\omega$ is not an independent variable but can be related to 
$x_1$ and $x_2$ through the gluon's on--shell condition 
$p_g^2=0$ and 4--momentum conservation $p_g=p_B-p_t-p_{\bar t}$. 
One obtains: 
\begin{equation}
0=1+2{m_t^2\over Q^2} -x_1-x_2+{1\over 2}x_1x_2(1-\beta_1\beta_2\cos\omega)
\label{dm2822}
\end{equation}
The 2--dimensional phase space integral is given by 
\begin{equation}
\text{PS}(B^\ast\rightarrow t\bar t g)={\pi^2Q^2\over 4} 
\int_0^{1-2m_t/Q} dy \int_{z_-}^{z_+} dz
\label{dm2823}
\end{equation}
where 
\begin{equation}
y:={2\bar t g\over Q^2}=1-x_1 \qquad 
z:={2 t g\over Q^2}=1-x_2
\label{dm2824}
\end{equation}
and 
\begin{equation}                                                        
z_\pm={1\over 2} {y\over y+{m^2\over Q^2}} 
[x_1(1\pm \beta_1)-2{m^2\over Q^2}]
\label{dm2825}
\end{equation}

The polarization vectors Eqs. (\ref{q5}) and (\ref{q52}) 
of the B--boson are not modified 
in higher orders, because $p_B$ has not changed. 
In contrast, the spinors for the fermions become more 
complicated than in lowest order:   
\begin{eqnarray} \nonumber
&\bar u_{+{1\over 2}} (p_t)=(a_{1+},0,a_{1-},0)
\quad
&\bar u_{-{1\over 2}} (p_t)=(0,a_{1-},0,a_{1+})
\\ \nonumber
& v_{+{1\over 2}} (p_{\bar t})=(a_{2+}s_2,-a_{2+}c_2,-a_{2-}s_2,a_{2-}c_2)
\quad
& v_{-{1\over 2}} (p_{\bar t})=(a_{2-}c_2,a_{2-}s_2,-a_{2+}c_2,-a_{2+}s_2)
\label{q6800}
\end{eqnarray}
where $s_2=\sin{\omega\over 2}$, $c_2=\cos{\omega\over 2}$ and 
$a_{i\pm}^2={Q\over 2}x_i(1\pm \beta_i)$, $i=1,2$.

Using these parametrizations, the real gluon amplitudes 
$M_I(B^\ast\rightarrow t\bar tg)(H,h,\bar h)$, $I=V,A$ and 
$h,\bar h=\pm{1\over 2}$, must be calculated. As in lowest 
order, it is sufficient to know the case $H=+1$.  
Before performing the phase space integrations, the 
density matrix must be formed. More precisely, we have 
calculated the normalized density matrices 
$R_{IJ}(+1,h,\bar h,+1,h',\bar h')$, $IJ=VV,VAAVAA$ and $AA$, 
Eq. (\ref{dm25}), 
corresponding to the finite combination Eq. (\ref{dm25aa}). 
Numerical results for all the 30 independent matrix elements 
are shown in Figures 2--61 as a function of the 
$e^+e^-$ energy $Q$ and for a top quark mass of $m_t=170$ GeV. 
The results are shown in the following 
form: each density matrix element has a decomposition of the 
form 
\begin{equation}
R_{IJ}(+1,h,\bar h,+1,h',\bar h')=C_1 {1\over 2}(1+c_t^2) 
+C_2s_t^2+C_3 c_t+C_4s_t+C_5c_ts_t
\label{dm28251}
\end{equation}
where $c_t=\cos\theta$, $s_t=\sin\theta$ and 
of course the coefficients $C_j$ depend 
on the spin quantum numbers $h,\bar h,h'$ and $\bar h'$. 
Each coefficient $C_j$  has a lowest order contribution,  
which however vanishes in many cases, and a higher order 
correction, i.e. 
\begin{equation}
C_j=\text{LO}+ \alpha  \, \, \text{HO}  
\label{dm282a2}
\end{equation}
where $\alpha=C_F{\alpha_s\over 2\pi}$ as before. 
In Figures 2--61 
there is always one figure displaying the lowest order term (LO) 
for each $j=1,2,3,4,5$ followed by the corresponding figure for the 
higher order term (HO); and these pairs of figures are 
successively presented for each combination $(+1,h,\bar h,+1,h',\bar h')$. 
Note that only the following combinations of 
$(h,\bar h,h',\bar h')$ are shown: 
$(----),(---+),(--+-),(--++),(-+-+),(-++-),(+-+-),(++-+),(+++-),(++++)$.
The rest follows from the symmetry of the density matrix, i.e. 
$R_{IJ}(+1,h,\bar h,+1,h',\bar h')=R_{IJ}(+1,h',\bar h',+1,h,\bar h)$. 
Some of the curves seem to increase rather strongly with increasing 
$Q^2$. However, it can be shown that all HO corrections converge 
to a constant at $Q\rightarrow \infty$. 
Corrections are typically at the percent level. This is not 
astonishing, because by forming the {\it normalized} density 
matrix, the bulk of the higher order correction due to the known 
overall K--factors \cite{report} drops out.
The odd figure numbers 3--61 really contain the main results 
of our paper, from which higher order QCD corrections to any 
angular distribution of top quark spins or top quark decay 
products can in principle be determined (for the latter one needs 
in addition the QCD corrections to the top quark decay 
amplitudes calculated in \cite{jirolampe}).

For convenience, we present in figures 62--91 the $\theta$ 
dependence of the matrix elements 
$R_{IJ}(+1,h,\bar h,+1,h',\bar h')$. More precisely, 
the matrix elements are written as 
$R_{IJ}(+1,h,\bar h,+1,h',\bar h')=\text{LO}+\alpha \, \, \text{HO}$ and 
$\text{LO}$ and $\text{HO}$ are displayed as a function of $\theta$ 
for a fixed total energy $Q=400$ GeV. 
Again, the corrections are relatively small because the normalized 
density matrix is considered, in which the correction 
to the trace of the density matrix drops out. 

\vskip1cm

{\bf \large 7. Summary}

In this article a complete decomposition into spin contributions 
of the production 
of top quarks in $e^+e^-$ annihilation has been given including 
higher order QCD corrections. 
The corrections are relatively small (on the percent level) 
because the {\it normalized}
density matrix was considered, in which the correction
to the trace of the density matrix drops out.
This means in turn that twoloop calculations are at most
necessary for the trace of the density matrix (corrections of up 
to 10\% \cite{report}), but not for any
of the normalized density matrix elements, because these $O(\alpha_s^2)$ 
corrections are most probably below the permille level.

Our results can be applied to 
calculate distributions for any decay product of top quarks 
at a future linear collider. I have not attempted to work out these 
applications, but have merely given in section 3 the  
necessary reduction formulae.  
To prove that the applications are rather straightforward, 
I would like to finish this work by making the QCD corrections 
to the $e^+e^-$ cross section Eq. (\ref{mgzd61}) explicit. 
Eq. (\ref{mgzd61}) can be rewritten as 
\begin{eqnarray}\nonumber
& & \sigma_\pm = \sum_{h,\bar h,h',\bar h'} D(h,\bar h) D^\ast(h',\bar h')
\\
& &
    \bigl\{ v_\pm^2 R_{VV}(\pm 1,h,\bar h,\pm 1,h',\bar h')
               (4+8{m_t^2\over Q^2}) (1+\alpha K_{VV}(\beta))
+4\beta^2(1+\alpha  K_{AA}(\beta))
 \\ & &
      [v_\pm a_\pm R_{VAAVAA}(\pm 1,h,\bar h,\pm 1,h',\bar h')
      +a_\pm (a_\pm-v_\pm)R_{AA}(\pm 1,h,\bar h,\pm 1,h',\bar h')  ]
      \bigr\}
\label{md610}
\end{eqnarray}
where $K_{VV}(\beta)$ and $K_{AA}(\beta)$ are the known corrections 
to the trace of the density matrix \cite{report}, cf. 
Eq. (\ref{dm251}). 
Note that $K_{VA}=K_{AV}=0$, so that there is an 
overall factor of $1+\alpha  K_{AA}$ in the second line of 
Eq. (\ref{md610}). Using Eq. (\ref{md610}) one can directly 
insert the corrections (figures 2--61) to obtain the cross 
sections including higher order QCD.

\vskip3cm





\end{document}